\input harvmac
 \noblackbox

\input epsf
\epsfverbosetrue
\def\epsfsize#1#2{\hsize}
\parindent =6pt

\def\B{{\cal B}}
\def\K{{\cal K}}

\def\p{^\prime}
\def\pp{^{\prime\prime}}


%
%
\def\RF#1#2{\if*#1\ref#1{#2.}\else#1\fi}
\def\NRF#1#2{\if*#1\nref#1{#2.}\fi}
\def\refdef#1#2#3{\def#1{*}\def#2{#3}}
\def\rdef#1#2#3#4#5{\refdef#1#2{#3, `#4', #5}}

%
%
\def\ts{\hskip .16667em\relax}

\def\FAP{{\it Funct.\ts Analy.\ts Appl.\ts}}

\def\JP{{\it J.\ts Phys.\ts}}

\def\NP{{\it Nucl.\ts Phys.\ts}}
\def\PL{{\it Phys.\ts Lett.\ts}}

\def\PR{{\it Phys.\ts Rev.\ts}}

\def\Zm{Zamolodchikov}
\def\AZm{A.B. \Zm}

\def\dur{H.\ts W.\ts Braden, E.\ts Corrigan, P.E. Dorey \ and R.\ts Sasaki}

%

%
%
\def\RF#1#2{\if*#1\ref#1{#2.}\else#1\fi}
\def\NRF#1#2{\if*#1\nref#1{#2.}\fi}
\def\refdef#1#2#3{\def#1{*}\def#2{#3}}
\def\rdef#1#2#3#4#5{\refdef#1#2{#3, `#4', #5}}

%
%
\def\ts{\hskip .16667em\relax}

\def\FAP{{\it Funct.\ts Analy.\ts Appl.\ts}}

\def\JP{{\it J.\ts Phys.\ts}}

\def\NP{{\it Nucl.\ts Phys.\ts}}
\def\PL{{\it Phys.\ts Lett.\ts}}

\def\PR{{\it Phys.\ts Rev.\ts}}

\def\Zm{Zamolodchikov}
\def\AZm{A.B. \Zm}

\def\dur{H.\ts W.\ts Braden, E.\ts Corrigan, P.E. Dorey \ and R.\ts Sasaki}

%
%
\refdef\rAFZa\AFZa{A.\ts E.\ts Arinshtein, V.\ts A.\ts Fateev and
 \AZm, \lq Quantum S-matrix of the 1+1 dimensional Toda chain',
 \PL {\bf B87} (1979) 389-392}

 \rdef\rAKLa\AKLa{M. Ameduri, R. Konik and A. LeClair}
 {Boundary sine-Gordon interactions at the free fermion point}
 {\PL {\bf B354} (1995) 376}

\rdef\rBa\Ba{P. Bowcock}
{Classical backgrounds and scattering for affine Toda theory on a half line}
{DTP-96-37; hep-th/9609233}

\rdef\rBCDRa\BCDRa{P. Bowcock, E. Corrigan, P.E. Dorey and R. H. Rietdijk}
{Classically integrable boundary conditions for affine Toda field theories}
{\NP {\bf B445} (1995) 469}

\rdef\rBCRa\BCRa{P. Bowcock, E. Corrigan and R. H. Rietdijk}
{Background field boundary conditions for affine Toda field theories}
{\NP {\bf B465} (1996) 350}

\rdef\rBCDSc\BCDSc{\dur}
{Affine Toda field theory and exact S-matrices}
{\NP {\bf B338} (1990) 689}

\refdef\rBCDSe\BCDSe{\dur,
\lq Multiple poles and other features of affine Toda field theory',
\NP {\bf B356} (1991) 469-98}

\rdef\rCf\Cf{E. Corrigan}
{Integrable field theory with boundary conditions}
{DTP-96/49; hep-th/9612138}

\rdef\rCDRa\CDRa{E.\ts Corrigan, P.E. Dorey, R.H.\ts Rietdijk}
{Aspects of affine  Toda field theory on a half line}
{Suppl. Prog. Theor. Phys. {\bf 118} (1995) 143}

\rdef\rCDRSa\CDRSa{E.\ts Corrigan, P.E. Dorey, R.H.\ts Rietdijk
and R.\ts Sasaki}
{Affine Toda field theory on a half line}
{\PL {\bf B333} (1994) 83}

\refdef\rCDSa\CDSa{E. Corrigan, P.E. Dorey and R. Sasaki,
 \lq On a generalised bootstrap principle',
\NP {\bf B408} (1993) 579-99}

\refdef\rDGZa\DGZa{G.W. Delius, M.T. Grisaru and D. Zanon,
\lq Exact S-matrices for non simply-laced affine Toda theories',
\NP {\bf B382}  (1992) 365-408}

\rdef\rDe\De{P.\ts E.\ts Dorey}
{A remark on the coupling dependence in affine Toda field theories}
{\PL {\bf B312} (1993) 291}

\rdef\rFKc\FKc{A.\ts Fring and R.\ts K\"oberle}
{Factorized scattering in the presence of reflecting boundaries}
{\NP {\bf B421} (1994) 159}

\rdef\rFKd\FKd{A.\ts Fring and R.\ts K\"oberle}
{Affine Toda field theory in the presence of reflecting boundaries}
{\NP {\bf B419} (1994) 647}

\rdef\rFSa\FSa{A. Fujii and R. Sasaki}
{Boundary effects in integrable field theory on a half line}
{Prog. Theor. Phys. {\bf 93} (1995) 1123}

\refdef\rGd\Gd{S.\ts Ghoshal,
`Boundary state boundary S-matrix of the sine-Gordon model',
{\it Int. J. Mod. Phys.} {\bf A9} (1994) 4801}

\refdef\rCKa\CKa{H.S. Cho and J.D. Kim, \lq Boundary reflection matrix for
$d_4^{(1)}$ affine Toda field theory', DTP-95-23; hep-th/9505138}

\rdef\rFSWa\FSWa{P. Fendley, H. Saleur and N.P. Warner}
{Exact solutions of a massless scalar field with a relevant boundary
interaction}
{\NP {\bf B430} (1994) 577}

\rdef\rIOZa\IOZa{T. Inami, S. Odake and Y-Z Zhang}
{Supersymmetric extension of the sine-Gordon theory with integrable
boundary interactions}
{\PL {\bf B359} (1995) 118}

\rdef\rKe\Ke{J.D. Kim}
{Boundary reflection matrix in perturbative quantum field theory}
{\PL {\bf B353} (1995) 213}

\rdef\rKf\Kf{J.D. Kim}
{Boundary reflection matrix for A-D-E affine Toda field theory}
{DTP-95-31; hep-th/9506031}

\rdef\rKg\Kg{J.D. Kim}
{Boundary reflection matrices for non simply-laced affine Toda
field theories}
{\PR {\bf D53} (1996) 4441}

\rdef\rKKa\KKa{J.D. Kim and I.G. Koh}
{Square root singularity in boundary reflection matrix}
{\PL {\bf B388} (1996) 550}

\rdef\rMv\Mv{A. MacIntyre}
{Integrable boundary conditions for classical sine-Gordon theory}
{{\it  J. Phys.} {\bf A28} (1995) 1089}

\rdef\rGZa\GZa{S.\ts Ghoshal and \AZm}
{Boundary $S$ matrix and boundary state in two-dimensional
integrable quantum
field theory}
{{\it Int. J. Mod. Phys.} {\bf A9} (1994) 3841}

\rdef\rMSa\MSa{M. Moriconi and K. Schoutens}
{Reflection matrices for integrable N=1 supersymmetric theories}
{\NP {\bf B487} (1997) 756}

\refdef\rPZa\PZa{S. Penati and D. Zanon,
\lq Quantum integrability in two-dimensional sytems with boundary',
\PL {\bf B358} 63}

\rdef\rSSWa\SSWa{H. Saleur, S. Skorik and N.P. Warner}
{The boundary sine-Gordon theory: classical and semi-classical analysis}
{\NP {\bf B441} (1995) 421}

\rdef\rSSa\SSa{S. Skorik and H. Saleur}
{Boundary bound states and boundary bootstrap in the sine-Gordon model
with Dirichlet boundary conditions}
{\JP {\bf A28} (1995) 6605}

\rdef\rSk\Sk{R.\ts Sasaki}
{Reflection bootstrap equations for Toda field theory}
{in {\it Interface between Physics and Mathematics}, eds
W. Nahm and J-M Shen,
(World Scientific 1994) 201}

\rdef\rSl\Sl{E.\ts K.\ts Sklyanin}
{Boundary conditions for integrable equations}
{\FAP {\bf 21} (1987) 164}

\rdef\rSm\Sm{E.\ts K.\ts Sklyanin}
{Boundary conditions for integrable quantum systems}
{\JP {\bf A21} (1988) 2375}
\rightline{DTP-97/33}
\rightline{hep-th/9707235}
\bigskip
{\obeylines \centerline{\bf On duality and  reflection factors for the
\ sinh-Gordon model }
\centerline{\bf with a boundary}}
\vskip 3pc
\centerline{E. Corrigan}
\bigskip
\centerline{\it Department of Mathematical Sciences}
\centerline{\it University of Durham}
\centerline{\it Durham DH1 3LE, England}
\medskip
\vskip 1pc

 \centerline {\bf Abstract}
 \bigskip
\noindent The sinh-Gordon model with  integrable boundary conditions
is considered in low order perturbation theory. It is pointed out
that results obtained by Ghoshal for the sine-Gordon breather
reflection factors
suggest an interesting dual
relationship between models with different boundary conditions.
Ghoshal's formula for the lightest breather is checked
perturbatively
to $O(\beta^2)$ in the special set of cases in which
the $\phi\rightarrow -\phi$ symmetry is maintained. It is noted that
the parametrisation of the boundary potential which is natural for
the
semi-classical approximation also provides a good parametrisation at the
`free-fermion' point.

\parindent=10pt

\newsec{Introduction}

In recent years, some progress has been made concerning the
question of integrability
 for two-dimensional field theories
restricted to a half-line\NRF\rGZa\GZa\NRF\rGd\Gd
\NRF\rFKc{\FKc\semi\FKd}
\NRF\rSk\Sk\NRF\rCDRa\CDRa\NRF\rCDRSa\CDRSa
\NRF\rBCDRa{\BCDRa\semi\BCRa}\NRF\rMv{\Sl\semi\Sm\semi\Mv}
\NRF\rSSWa{\SSWa\semi\SSa}\NRF\rPZa\PZa
\hskip -4pt
\refs{\rGZa -\rPZa}, or to
an interval. (For a recent review of some aspects of this,
see \NRF\rCf\Cf\refs{\rCf}.)
In this article
it is intended to concentrate on the simplest example, other
than free field theory, and to estimate using perturbation
theory the reflection
factors corresponding to those boundary conditions which
preserve integrability. Thus, the model to be studied is the
sinh-Gordon model restricted to the region $x<0$. There are
several reasons for doing this. Firstly, Ghoshal \refs{\rGd}
has suggested
a formula for the reflection factors for the breather states of
the sine-Gordon model using the reflection bootstrap equations
and the soliton reflection factors suggested earlier
by Ghoshal and Zamolodchikov \refs{\rGZa}. It is plausible to
suppose that
the reflection factor for the lightest breather state, analytically
continued to imaginary coupling, should provide a sensible
hypothesis for the reflection factors for the sinh-Gordon particle.
On the other hand, the reflection factors of the sinh-Gordon
model are accessible perturbatively and so a start can be made
towards checking Ghoshal's formula. Unfortunately, the relationship
between the two classical parameters which may be introduced via
the boundary potential and the quantum reflection factors
is unclear except in certain cases. Therefore, the perturbative
calculation is potentially a useful way to suggest the missing
connection. In those cases where the relationship between classical
and quantum parameters is already known, perturbation theory
should provide a verification. The second reason concerns
weak-strong coupling duality.

The S-matrix describing the elastic
scattering of a pair of sinh-Gordon particles is known
\NRF\rAFZa\AFZa\refs{\rAFZa} to have the form
\eqn\sgsmatrix{S(\Theta )=-\, {1\over (B) (2-B)}}
where
\eqn\Bdef{B={1\over 2\pi}\ {\beta^2\over 1 + \beta^2 /4\pi}}
and
\eqn\blockdef{(x)={\sinh\left({\Theta\over 2}+{i\pi x\over 4}
\right)\over\sinh\left({\Theta\over 2}-{i\pi x\over 4}
\right)}.}
The rapidity difference of the two particles is represented by
$\Theta$.
This is clearly invariant under the transformation
\eqn\duality{\beta \rightarrow {4\pi\over\beta},}
and this invariance is referred to as the weak-strong coupling
duality. This form of duality is shared by all the other affine
Toda models based on the $ade$ series of data
\NRF\rAFZa\AFZa\NRF\rBCDSc{\BCDSc\semi\BCDSe}\refs{\rAFZa ,\rBCDSc},
and a generalised
form of it is enjoyed by those models based on the data
derived from the other (ie not
simply-laced) affine diagrams \NRF\rDGZa\DGZa\NRF\rCDSa{\CDSa
\semi\De}\refs{\rDGZa ,
\rCDSa}. In some earlier works concerning the
reflection factors \refs{\rFKc ,\rSk} it was suggested that they should
enjoy the same duality property. However, this is unlikely to
be the case, as the following argument demonstrates.

When there is a boundary condition at $x=0$, provided it is chosen
to maintain integrability, then it is expected that a single particle
approaching the boundary will be elastically reflected from it so
that
$$|\theta >_{\rm out}=\K (\theta )|-\theta >_{\rm in},$$
where $\theta$ is the rapidity of the particle. In the sinh-Gordon
model there is only one such particle.
Ghoshal's formula \refs{\rGd} for the sinh-Gordon reflection factors, after
translation to the above notation, has the form:
\eqn\ghoshal{\K(\theta )={(1)\, (2-B/2)\, (1+B/2)\over
(1-E(\beta ))\, (1+E(\beta ))\, (1-F(\beta ))\, (1+F(\beta ))},}
where $E$ and $F$ are two functions of $\beta$ which also depend on
the boundary parameters, but which are not yet fully determined
in terms of those parameters.\foot{In Ghoshal's notation $E=B\eta/\pi,
\ F=iB\vartheta/\pi$.} However, for the Neumann boundary
condition Ghoshal requires
\eqn\neumann{F=0,\qquad E= 1-B/2}
leading to a reflection factor $\K_N$, given by:
\eqn\neumannK{\K_N ={(1+B/2)\over (B/2)\, (1)}.}
Clearly, this is not self-dual; rather its dual partner is $\K^*_N$,
given by:
\eqn\neumannKdual{\K^*_N={(2-B/2)\over (1-B/2)\, (1)}.}

What about a verification of \neumann\ perturbatively? In effect
this has been provided implicitly  by
Kim \NRF\rKe\Ke\NRF\rKf{\Kf\semi
\Kg\semi\KKa}\refs{\rKe ,\rKf} who has developed perturbation
theory for affine Toda models satisfying the Neumann boundary
condition, and has provided detailed calculations up to $O(\beta^2)$.
In other words, the $O(\beta^2)$ check can be made merely by expanding
\neumannK\ and comparing with the formulae given by Kim \refs{\rKe}.
Explicitly,
from \neumannK ,
\eqn\kimcompare{\K_N\, \sim\, 1-{i\beta^2\sinh\theta\over 8}
\left[{1\over \cosh\theta -1}-{1\over\cosh\theta}\right],}
which is in perfect agreement.\foot{Kim did not comment
on this agreement.
He chose to assume instead that the $a_1^{(1)}$ reflection factor
should be
self-dual and proposed
an alternative formula for $\K_N$, disagreeing with Ghoshal, but
nevertheless in agreement
with perturbation theory to $O(\beta^2)$.}

To discover which boundary condition is dual to the Neumann condition
it will be enough to consider the classical limit of \neumannKdual .
However, first it is necessary to consider the classical limit
of \ghoshal . That is, taking $\beta\rightarrow 0$,
\eqn\ghoshalclassK{\K_0(\theta )=-\, {(1)^2\over
(1-E(0))\, (1+E(0))\, (1-F(0))\, (1+F(0))}.}
Fortunately, this may be calculated independently and the
result was already reported some time ago \refs{\rCDRa}.

In order to state the result of the
`classical' calculation  further details
of the model are required. Its equation of motion and boundary
condition are:
\eqn\sgeq{\eqalign{\partial^2\phi &= -{\sqrt{2}\over\beta}
\left(e^{\sqrt{2}
\beta\phi}-e^{-\sqrt{2}\beta\phi}\right)\qquad\ \ \qquad x<0\cr
{\partial\phi\over \partial x}&=-{\sqrt{2}\over\beta}\left(\sigma_1e^{
\beta\phi /\sqrt{2}}-\sigma_0e^{-\beta\phi /\sqrt{2}}\right)\qquad
x=0,\cr}}
where the $\sqrt{2}$ is a conventional normalisation in Toda theory.
The two constants $\sigma_0$ and $\sigma_1$ are essentially free
and represent the degrees of freedom permitted at the boundary \refs{\rGZa}.
The
constraints on the boundary
parameters are discussed in \refs{\rCDRa} and by Fujii and Sasaki
\NRF\rFSa\FSa\refs{\rFSa}. The
boundary condition is required to have the given form as a consequence
of maintaining integrability on the half-line. It is also convenient to
write
\eqn\defepsilon{\sigma_i=\cos a_i\pi .}
Then, the classical reflection factor is determined by first finding the
lowest energy static solution to \sgeq\ (the `vacuum configuration')
and then solving the linearised scattering problem in this static
background. Further details of this are given below and the result for
the reflection factor calculated in this approximation (and for
$|a_i|\le 1$) is:
\eqn\classK{\K_0(\theta )=-\, {(1)^2\over
(1-a_0-a_1)\, (1+a_0+a_1)\, (1-a_0+a_1)\, (1+a_0-a_1)}.}

The formula \classK\ clearly has the same form as the classical
limit of Ghoshal's
formula, and is unity when $a_0=a_1=1/2$ (the Neumann condition),
in agreement with the classical limit of $\K_N$, eq\neumannK .

On the other hand, the classical limit of $\K^*_N$ is not unity.
Rather, it is $-1/(1)^2$, and is in agreement with \classK\ for the
choice $a_0=a_1=0$. In other words, it is tempting
to conclude that the two boundary conditions
\eqn\dualpair{{\partial\phi\over \partial x}=0 \qquad \hbox{and}
\qquad {\partial\phi\over \partial x}=-{\sqrt{2}\over\beta}\left(e^{
\beta\phi /\sqrt{2}}-e^{-\beta\phi /\sqrt{2}}\right)\qquad
\ \hbox{at}\quad x=0,}
are dual, in the sense that their reflection factors are transformed into
each other under the transformation \duality .

It should now be clear that it is the functional dependence of $E$ and
$F$ on the three parameters $a_0$, $a_1$ and $\beta$ which must
be found together with a proper understanding of which pairs of
boundary data are dual to each other, and which (if any) are
self-dual. The rest of the paper is concerned with these questions.

\newsec{The propagator in a general background}

To calculate the propagator it will be necessary to consider the
linear perturbation around the static background solution to
\sgeq . The static solutions have the form
\eqn\background{e^{\beta\phi_0 /\sqrt{2}}={1+e^{2(x-x_0)}\over
1-e^{2(x-x_0)}}}
where the parameter $x_0$ is determined by the boundary condition
and is given by:
\eqn\constant{\coth x_0 =\sqrt{1+\sigma_0\over 1+\sigma_1}.}
Note that provided $\sigma_0 \ge \sigma_1$ it is
guaranteed that  $x_0\ge 0$; if on the other hand
$\sigma_1 \ge \sigma_0$ it would be necessary to take the
background provided by  the inverse of the right hand side of
\background . In the calculations which follow it will be assumed
$\sigma_0 \ge \sigma_1$ for definiteness.

Linearising \sgeq\ around the static background leads to a pair of
equations for the first order correction to $\phi$:
\eqn\linear{\eqalign{\partial^2\phi_1 +4\left(1+{2\over \sinh^2 2(x-x_0)}
\right)\phi_1&=0\qquad x<0\cr
{\partial\phi_1\over \partial x} +\left(\sigma_1\sqrt{1+\sigma_0
\over 1+\sigma_1}
+\sigma_0\sqrt{1+\sigma_1\over 1+\sigma_0}\right)\phi_1&=0
\qquad x=0.\cr}}
 Fortunately, these equations
 may be solved exactly \refs{\rCDRa} and the required Green's
function can be written down explicitly once the eigenfunctions of the
second order differential operator in the first of eqs\linear\
have been determined. It is convenient to write $\phi_{k,\omega}$
to denote the eigenfunction corresponding to the eigenvalue
$\omega^2-k^2-4$, and then
\eqn\eigenfunction{\phi_{k,\omega}=ie^{-i\omega t}r(k)\left(
F(k,x)e^{ikx}+F(-k,x)e^{-ikx}\right) ,}
where $r(k)$ is a real, even function of $k$ (which will be determined
by normalising the Green's function properly), and where $F(k,x)$ is
given by:
$$F(k,x)=P(k)\left(ik-2\coth 2(x-x_0)\right),$$
with
$$P(k)=(ik)^2-2ik\sqrt{1+\sigma_0}\sqrt{1+\sigma_1}+
2(\sigma_0+\sigma_1).$$
The classical reflection factor \classK\ follows immediately
from \eigenfunction\ on setting $\omega =2\cosh\theta,\ k=2\sinh\theta$,
calculating $F(-k,-\infty)/F(k,-\infty)$, and using the
convenient parametrisation \defepsilon .

The Green's function, or configuration space propagator,
constructed from these eigenfunctions is not
hard to find. It is:
\eqn\green{G(x,t;x^\prime ,t^\prime )=i \int {d\omega\over 2\pi}
 \int {dk\over 2\pi}\ {\cal G}(\omega,k;x,t;x^\prime ,t^\prime )}
 where
 \eqn\kernel{{\cal G}(\omega,k;x,t;x^\prime ,t^\prime )=e^{-iw(t-t^\prime)}
 {f(k,x)f(-k,x^\prime )e^{ik(x-x^\prime)}+\K_0f(-k,x)f(-k,x^\prime )
e^{-ik(x+x^\prime)}\over \omega^2-k^2-4 +i\epsilon}}
$$f(k,x)={ik-2\coth 2(x-x_0)\over ik+2},$$
and
$$\K_0={P(-k)\over P(k)}\ {ik-2\over ik+2}.$$
When $k=2\sinh\theta$ this expression for $\K_0$ reduces
to the previous one, eq\classK .
To check that \green\ is normalised correctly it is enough to note
that it is in the limit $x,x^\prime \rightarrow -\infty$.

If $\sigma_0=\sigma_1=\sigma$, eq\green\ simplifies because in that case
$f(k,x)=1$ and the factor $\K_0$ reduces to
$$\K_\sigma= {ik+2\sigma\over ik-2\sigma}.$$
In other words, in this special set of cases,
\eqn\specialkernel{{\cal G}(\omega ,k;x,t;x^\prime ,t^\prime )=
e^{-iw(t-t^\prime)}\
 {e^{ik(x-x^\prime)}+\K_\sigma
e^{-ik(x+x^\prime)}\over \omega^2-k^2-4 +i\epsilon}}
In this situation, the static background is simply $\phi_0=0.$

In this article a pragmatic approach will be taken
\NRF\rKe\Ke\refs{\rKe}.
The classical
reflection factor is an integral part of the free field propagator
calculated within the classical background. The full two-point
function may then be calculated perturbatively (once subtractions
are made to remove infinities), and provides a working definition of
the quantum reflection factor in the sense that it will be the
coefficient
of $e^{-ik(x+x\p )}$ in the residue of the on-shell pole in the
asymptotic region $x,\ x\p\rightarrow -\infty$.

\newsec{Low order perturbation theory}

Since perturbation theory will be developed around an $x$-dependent
background field configuration, it is possible to use a more-or-less
standard Feynman expansion in configuration space, using the above
propagator
but recognising that the `vertices' will be position dependent.
One consequence
of this  is that although the sinh-Gordon theory on the whole line has
only vertices at which an even number of lines join, the theory
on the half-line  must also, in general,  contain vertices at
which an odd number of lines meet. In other words, the symmetry
$\phi\rightarrow -\phi$, enjoyed by the bulk theory, is broken by
most boundary conditions. The exceptions to this are precisely
the boundary conditions for which $\sigma_0=\sigma_1$. In those
cases, the classical background is $\phi_0=0$ and there will be no
additional vertices compared with  the bulk
theory other than those arising specifically from the boundary term
itself.
Moreover, the boundary terms  will only generate
vertices  at which an even number of lines meet. In this article,
it is intended only to perform detailed calculations in the special
case $\sigma_0=\sigma_1=\sigma$. The more general situation will be
treated elsewhere since the computations are more lengthy and intricate.

To $O(\beta^2)$, there are two contributions which need to be calculated
and each of them
is a single loop diagram. The first comes from the boundary and is
described by:
\eqn\boundaryloop{\B_1 (x,t;x^\prime ,t^\prime )= -i \sigma\beta^2 C_1
\int_{-\infty}^\infty dt\pp G(x,t;0,t\pp )G(0,t\pp ;0,t\pp )
G(0,t\pp ;x\p,t\p ),}
where $C_1$ is a combinatorial factor. The second comes from the bulk
potential and has the form
\eqn\bulkloop{\B_2 (x,t;x^\prime ,t^\prime )=-8 i\beta^2 C_2
\int_{-\infty}^\infty dt\pp \int_{\infty}^0 dx\pp G(x,t;x\pp ,t\pp )
G(x\pp ,t\pp ;x\pp ,t\pp )
G(x\pp ,t\pp ;x\p,t\p ),}
where $C_2$ is another combinatorial factor. The relative factor of $8$
between the two expressions arises from the different factors of
$\sqrt{2}$ in the bulk and boundary potential terms. The expression
\boundaryloop\ is relatively simple to analyse and it will be
considered first.

The integral over $t\pp $ in expression \boundaryloop\ gives a delta
function allowing the $\omega\p$, say, integral to be done leaving
the momentum integrals for all three propagators, and the frequency
integral for the loop (middle propagator) to be performed.

The middle propagator is clearly divergent and a minimal subtraction
will be made to yield a finite part. That is, the logarithmically
divergent integral
$$\int{d\omega\pp\over 2\pi}\int{dk\pp\over 2\pi}{i\over {\omega\pp}^2
-{k\pp }^2 -4+i\epsilon}\left(1 + {ik\pp +2\sigma\over ik\pp -2\sigma}
\right)$$
will be replaced by the finite part
$$\int{d\omega\pp\over 2\pi}\int{dk\pp\over 2\pi}{i\over {\omega\pp }^2
-{k\pp }^2 -4+i\epsilon}\left({4\sigma\over ik\pp -2\sigma}\right)$$
which can be evaluated to obtain first
$${1\over 2} \int{dk\pp\over 2\pi}{1\over\sqrt{{k\pp}^2+4}}\,
\left({4\sigma\over ik\pp -2\sigma}\right) ,$$
and then finally, using \defepsilon , to obtain
\eqn\middle{-(a\ \hbox{mod}\ 1) \ {\cos a\pi\over \sin a\pi}.}
The expression \middle\  is invariant under the transformation
$a\rightarrow a+1$.

The momenta of the other two propagators may be integrated out by
closing the contours in the upper half plane (first setting
$k\rightarrow -k$ in the first term of the first propagator).
Provided $\sigma>0$,
the reflection factor term does not contribute a pole. On the
other hand, if $\sigma<0$, there is an extra pole but its
contribution may be discounted in the limit $x,x\p\rightarrow -\infty$,
because it will be exponentially damped. Thus,
the two integrals reduce to
$$\int{dk\over 2\pi}\int{dk\p\over 2\pi}{i\over\omega^2-
{k\phantom{\p}}^2-4+i\epsilon}\
{i\over\omega^2-{k\p}^2-4+i\epsilon}e^{-i(kx+k\p x\p)}\left({2ik\over
ik-2\sigma}\right)\left({2ik\p\over
ik\p-2\sigma}\right)$$
and hence to,
\eqn\rest{\left(-{i\over 2}\right)^2 \left({1\over \widehat{k}}\right)^2
e^{-i\widehat{k}(x+x\p )}\left({2i\widehat{k}\over i\widehat{k}-2\sigma}
\right)^2,}
where $\widehat{k}=\sqrt{\omega^2 -4}$.

Combining the results contained in \middle\ and \rest , the contribution
\boundaryloop\ becomes
\eqn\boundarypiece{\B_1 ={i a \beta^2\over 8\tan^2 a\pi} C_1\int
{d\omega\over 2\pi}\, \K_0(\widehat{k})\, e^{-i\omega (t-t\p)}
e^{-i\widehat{k}(x+x\p )} \left({1\over \cosh\theta -\sin a\pi}-
{1\over\cosh\theta +\sin a\pi}\right),}
where the integral over $\omega$ is to be understood as a positive
energy integral, and it is convenient to set $\widehat{k}=2\sinh\theta$
in the last terms of the integrand. The overall factor of $a$ should be
interpreted modulo unity. Summarising, the correction to
the reflection factor from the boundary piece is
\eqn\correctboundary{{ia\beta^2 C_1\over 2 \tan^2 a\pi} \sinh\theta
\left({1\over \cosh\theta -\sin a\pi}-
{1\over\cosh\theta +\sin a\pi}\right).}

The evaluation of \bulkloop\ proves to be more testing. The integral over
$t\pp $ is straightforward again and provides a delta function allowing
one of the non-loop energy  integrals ($\omega$ or $\omega\p$) to be
performed immediately. The loop integral is
$$ \int{d\omega\pp\over 2\pi}{dk\pp\over 2\pi}
\ {i\over {\omega\pp}^2-{k\pp}^2-4+i\epsilon}
\ \left( 1+\K_\sigma (k\pp )\ e^{-2ik\pp x\pp}\right),$$
which is logarithmically divergent. However, a minimal subtraction of the
term without the exponential dependence on a position coordinate leaves
a finite integral. This subtraction is natural, coinciding with what
would be achieved by normal-ordering in the usual bulk theory on
the full line, as has been emphasised by Kim \refs{\rKe}. What remains
is:
\eqn\bulkmess{\eqalign{\int_{-\infty}^0 dx\pp&\int{d\omega\over 2\pi}
{dk\over 2\pi} \ e^{-i\omega (t-t\p )}\ {i\over \omega^2-k^2-4+i\epsilon}
\ \left(e^{-ik(x-x\pp )}+\K_\sigma (k)e^{-ik(x+x\pp )}\right)\cr
&\int{d\omega\pp\over 2\pi}{dk\pp\over 2\pi}
\ {i\over {\omega\pp}^2-{k\pp}^2-4+i\epsilon}
\ \K_\sigma (k\pp )\ e^{-2ik\pp x\pp}\cr
&\int{d\omega\p\over 2\pi}{dk\p\over 2\pi}
\ {i\over {\omega\p}^2-{k\p}^2-4+i\epsilon}
\ \left(e^{ik\p (x\pp -x\p )}+\K_\sigma (k\p )
e^{-ik\p (x\p +x\pp )}\right) .\cr}}
The integration over $x\pp$ is achieved using the device
$$\int_{-\infty}^0 dx\pp\ e^{(i\lambda +\rho ) x\pp}=
- \ {i\over \lambda -i\rho}\ ,
$$
where $\rho$ is a positive constant which will be taken to
zero at the end of the calculation.
This introduces a collection of poles for
the $k\pp$ integration which need to be dealt with separately.

The pole contributions after integrating out $\omega\pp$ are
$$\eqalign{\ \ -{i\over 2}\int&{d\omega\over 2\pi}{dk\over 2\pi}
{dk\p\over 2\pi}
\ e^{-i\omega (t-t\p )}e^{-i(kx-k\p x\p )}{i\over
{\omega\phantom{\p}}^2-{k\phantom{\p}}^2-4+i\epsilon}
\ {i\over {\omega\p}^2-{k\p}^2-4+i\epsilon}\cr
&\int{dk\pp\over 2\pi}\ {1\over \sqrt{{k\pp}^2+4} }\ \K_\sigma (k\pp )\
\left[{1\over k-2k\pp +k\p -i\rho} +{\K_\sigma (k\p )
\over k-2k\pp -k\p -i\rho}\right.\cr
&\ \ \ \ \ \ \ \ \ \ \ \ \ \ \ \ \ \ \ \ + \left. {\K_\sigma (k )
\over -k-2k\pp +k\p -i\rho}
+ {\K_\sigma (k )\K_\sigma (k\p )\over
-k-2k\pp -k\p -i\rho}\right] , \cr}$$
and the $k\pp$ integrations may be done by collapsing the contour into
the upper half-plane and onto the branch cut running from
$k\pp=2i$ to infinity along the imaginary axis, avoiding all the poles
(the `$i\rho$' effectively indicates the sense in
which the $k\pp$ contours must be interpreted), with the possible
exception of the pole in $\K_\sigma (k\pp )$.
If $\sigma >0$ this pole is avoided automatically; if $\sigma <0$,
it cannot be avoided but its residue integrated over $k$ and $k\p$
yields exponentially decreasing terms as $x,x\p\rightarrow -\infty$.

The integrals along the cut are all reducible to integrals of the
following form
\eqn\fancy{\int_2^\infty dy\ {1\over \sqrt{y^2-4\phantom{^2}}}\
{1\over y+ 2\zeta}
= {1\over\sqrt{1-\zeta^2}}\ \left({\pi\over 2}-
\tan^{-1}\sqrt{1+\zeta\over 1-\zeta}\right) .}
Thus, for example, with a little manipulation,
$$\eqalign{\int{dk\pp\over 2\pi}\ {1\over \sqrt{{k\pp}^2+4} }\
\K_\sigma (k\pp )&
{1\over k-2k\pp +k\p -i\rho}={1\over \pi} \int_2^\infty dy
{1\over \sqrt{y^2-4\phantom{^2}}}\ \K_\sigma (iy)\ {1\over k+k\p -2iy}\cr
&={i\over 2\pi}\int_2^\infty dy
{1\over \sqrt{y^2-4\phantom{^2}}}\left({\K_\sigma ({k+k\p\over 2})\over
y+i(k+k\p )/ 2}+{1-\K_\sigma ({k+k\p\over 2})\over y+2\sigma}\right),\cr
}$$
which can be evaluated using \fancy , to yield:
\eqn\kppbits{\eqalign{{i\over 2\pi}\left[{2\K_\sigma ({k+k\p\over 2})\over
\sqrt{4+(k+k\p )^2 /4}}\left( {\pi\over 2}\right.\right.&
\left.\left.-\tan^{-1}\sqrt{{4+i(k+k\p )\over
4-i(k+k\p )}}\right) \right.\cr
& \left.  + {1-\K_\sigma ({k+k\p\over 2})\over
\sqrt{1-\sigma^2} }
\left({\pi\over 2}-\tan^{-1}\sqrt{1+\sigma\over 1-\sigma}\right)
\right] .\cr}}
The $k,k\p$ integrals can be performed next and, up to vanishingly
small factors as before, one obtains as far as \kppbits\ is concerned
a relatively simple result,
\eqn\kppbitsa{{i\over 2\pi}\left[{2\K_\sigma (\widehat{k})\over
\sqrt{4+\widehat{k}^2}}\left( {\pi\over 2} -i\theta \right)+
\left( 1-\K_\sigma (\widehat{k})\right){a\pi\over 2\sin a\pi}\right].}
The other three pole pieces may be treated in the same manner yielding
contributions similar to \kppbits , except that $k+k\p$ is replaced by
one of $k-k\p$, $-k+k\p$ and $-k-k\p$, in turn, leading to corresponding
differences in the analogues of \kppbits . Assembling all four
contributions, and including the numerical factors
from the integrations leads to the following correction
to the reflection factor:
\eqn\correctbulk{\eqalign{-{i\beta^2 C_2\over 2} \K_\sigma (\widehat{k})
\sinh\theta\Biggl[&{1\over 2}\left({1\over
\cosh\theta +1} -{1\over \cosh\theta}\right)\cr
& +{ a\over\  \sin^2a\pi}\left(
{1\over \cosh\theta -\sin a\pi}-{1\over \cosh\theta +\sin a\pi}
\right)\Biggr].\cr} }
In \correctbulk , the overall factor $a$ is restricted to the range
$0\le a\le 1/2$, in order that $\sigma$ be positive.

The  combinatorial factors $C_1$ and $C_2$ are equal since they
arise from four-point interactions which are in essence identical,
differing only by constant factors.
Moreover, the common factor is $1/2$, as is easily checked using,
for example, the functional integral form of the Green's function.

Finally, notice that the two contributions from the boundary summarised in
\correctboundary\  and \correctbulk\ combine neatly (within the
restricted $a$
range)
to give the
result:
\eqn\Kcorrection{\eqalign{\delta\K_\sigma (\widehat{k})=
-{i\beta^2 \over 8} \K_\sigma (\widehat{k})
\sinh\theta\Biggl[&\left({1\over
\cosh\theta +1} -{1\over \cosh\theta}\right)\cr
& + 2a\left(
{1\over \cosh\theta -\sin a\pi}-{1\over \cosh\theta +\sin a\pi}
\right)\Biggr].\cr} }
For the special case $a=1/2$, \Kcorrection\ collapses to \kimcompare ,
as expected.

\newsec{Comparison with Ghoshal's formula}

To make the comparison with Ghoshal's formula up to $O(\beta^2)$
the expansion of \ghoshal\ is required. It is:
\eqn\ghosgalexp{\K (\theta) \sim \K_0(\theta )\left(1-{i\beta^2\over 8}
\sinh\theta \ G(\theta )\right),}
where
\eqn\G{\eqalign{G(\theta )=\ &{1\over \cosh\theta +1}-{1\over
\cosh\theta}\cr
&\ \ \ +{e_1\over \cosh\theta +\sin (e_0\pi / 2)}
-{e_1\over \cosh\theta - \sin  (e_0\pi / 2)}\cr
&\ \ \ +{f_1\over \cosh\theta +\sin  (f_0\pi / 2)}
-{f_1\over \cosh\theta -\sin (f_0\pi / 2)}\, ,\cr}}
and $E\sim e_0+e_1\beta^2/4\pi ,\ F\sim f_0+f_1\beta^2/4\pi $. Comparison
with the classical reflection factor $\K_0 (\theta )$, eq\classK\
reveals that $e_0 =a_0+a_1,\ f_0=a_0-a_1$; comparison with eq\Kcorrection\
reveals further that when $a_0=a_1=a$,
$$E\sim a\left(2-{\beta^2\over 2\pi}\right).$$
However, since $f_0$ vanishes, nothing can be concluded to
this order concerning $F$. Nevertheless, it is satisfying that the
correction
\Kcorrection\ is of the same general form as \G .

Given the coupling constant always appears to
enter in the combination $B(\beta )$, it is tempting to suppose on the
basis of the above that
\eqn\Efull{E(a,\beta )=2a \left(1-{B\over 2}\right).}

On the other hand, consider the weak-strong coupling transformation
applied to \ghoshal . The numerator  of \ghoshal\ is invariant under
a duality transformation and,  if it is assumed there is
another theory within the same class whose boundary condition
implies the dual reflection factor, there must be an $a^*$ such that
\eqn\alternatives{\hbox{\bf either}\ \ E(a, 4\pi /\beta ) =
\pm E( a^*,\beta )\ \hbox{mod}\ 4\ \
\hbox{\bf or}\ \ E(a, 4\pi /\beta ) = \pm F( a^*,\beta )\ \hbox{mod}\ 4\ .}
This is a plausible assumption given the invariance of the two-particle
S-matrix under the duality transformation, and given each boundary
condition within the special class $a_0=a_1$ preserves the symmetry
$\phi\rightarrow -\phi$. The first possibility within \alternatives\
cannot work given the form of \Efull  ---there is simply no
suitable $a^*$. On the other hand, the second alternative is possible
and suggests
\eqn\Ffull{F(a,\beta )=(1-2a){B\over 2}\ \ \hbox{and}\ \
a^*={1\over 2}-a.}
With $F$ given by \Ffull , there would be no change to the
perturbation theory result \Kcorrection\
up to the order given. On the other hand, the next order
should provide a test of the duality argument.

However, the correct formulae cannot be so simple.
Recall that a Dirichlet condition $\phi (0,t)=0$ also belongs to the special
class, on allowing $a$ to be pure imaginary and infinite. As before,
the perturbative
result  may be obtained from \G . On taking the appropriate
limit, all that remains is the first part, in perfect agreement with
Ghoshal's formula for the Dirichlet case, which  is self-dual and
given by
$$\K_D= {(2-B/2)(1+B/2)\over (1)}.$$
Within the special cases, it would be obtained by taking
$E\rightarrow i\infty$
and $F\rightarrow 0\ \hbox{mod\ 4}$. Clearly, the correct formula
would not be obtained
using \Efull\ and \Ffull : \Efull\ is satisfactory  but \Ffull\ is not.

Moreover, the special coupling constant for which $B=-2$, the
`free-fermion' point of the sine-Gordon model,\foot{It is also a special
point of the sinh-Gordon model in the sense that the S-matrix
\sgsmatrix\ is unity there.} has been analysed before
\NRF\rAKLa\AKLa\refs{\rAKLa}
and relations have been written down between the
parameters $\eta$ and $\vartheta$ occurring in Ghoshal's formula
at this value.
Explicitly, using the notation of \refs{\rAKLa} but in the normalisations
of this paper,
the ratio of the boundary to bulk coupling $\gamma$, and the parameter
$\phi_0$
are given by
$$\gamma = 4 \cos a_0\pi \cos a_1\pi , \qquad\cosh\phi_0 ={1\over 2}
\left({\cos a_0\pi\over \cos a_1\pi}+
{\cos a_1\pi\over \cos a_0\pi}\right).$$
 In terms of these,
$$ k=(1-\gamma\cosh\phi_0 +\gamma^2/4)^{-1/2}, \qquad \cos\xi =
k(1-\gamma\cosh\phi_0 /2),$$
and the two equations \refs{\rGZa},
$$\cos\eta \cos i\vartheta = -(\cos\xi )/k, \qquad
\cos^2\eta +\cos^2 i\vartheta =1 +1/k^2, $$
are solved by
\eqn\results{\cos\eta =\pm\cos (a_0 +a_1)\pi , \qquad \cos\vartheta =
\pm\cos (a_1-a_0)\pi .}
Hence, it is simple to deduce that at the free fermion point, and within
the cases $a_0=a_1$,
$${B\eta\over \pi}=E=4a\qquad {iB\vartheta\over\pi}=F=0\ \ \ \
\hbox{mod}\ 4.$$
Again, the formula \Efull\ is adequate but the formula \Ffull\ is
not.

A more general analysis might proceed along the following lines. Instead
of \Efull\ and \Ffull , one could write $(q=\beta^2/4\pi$)
\eqn\newtry{E(a,q)=2 {e(a,q)\over 1+q}\qquad F(a,q)=2{q f(a,q)\over 1+q},}
in which case, the duality transformation would require,
$$e(a,1/q)=\pm f(a^*,q)\ \hbox{mod}\ 2\left({1+q\over q}\right)
\qquad f(a,1/q)=
\pm e(a^*,q)\ \hbox{mod}\ 2(1+q).$$
Then, taking $e(a,q)\equiv a$  would be a consistent choice
(as suggested by the known facts
concerning $e(a,q)$), provided
\eqn\frelation{a=\epsilon_1f\bigl(\epsilon_2f(a,1/q)\ \hbox{mod}\
2(1+q), q\bigr)
\ \hbox{mod}\ 2\left({1+q\over q}\right) ,}
where $\epsilon_1$ and $\epsilon_2$ are choices of sign. However, it
has not yet proved possible to obtain a satisfactory conjecture based on
this idea. For example, it
is straightforward to see that choices of $f(a,q)$ of the form
$$f(a,q)={A(q)+aB(q)\over C(q)+aD(q)}$$
are inadequate once all the constraints are taken into account.

Alternatively, one might return to the first of \alternatives\ and seek
a formula for $E(a,q)$ of the form
\eqn\finaltry{E(a,q) ={n_0(a)+n_1(a)q+n_2(a)q^2 +\cdots +n_k(a)q^k\over
d_0(a)+d_1(a)q+d_2(a)q^2 +\cdots +d_k(a)q^k},}
where the coefficients of the polynomials are constrained by the data
described above and by the duality requirement. Neither linear nor
quadratic polynomials appear to be satisfactory, while cubics and
polynomials of higher degree are insufficiently constrained.

\newsec{Discussion}

Clearly, there is much left undone. Ghoshal's formula has been verified
to lowest order and it has proved possible to find a compact parametrisation
for the free-fermion point. However, it has not yet allowed a consistent
conjecture for the general parameter dependence even within the special class
of models considered here.
In the general case, $a_0\ne a_1$, the calculations are more intricate
for the reasons stated before.

It is intriguing to wonder about the supersymmetric
sinh-Gordon theory. It has been noted
\NRF\rIOZa{\IOZa\semi\MSa}\refs{\rIOZa}
that only a few boundary conditions are compatible with both
supersymmetry and integrability. Moreover, they belong to the special
class considered here, at least as far as the bosonic part is concerned
(and in the normalisation of this paper correspond to
$\sigma_0=\sigma_1= \pm 1$). It would be interesting to understand
the effect of the fermions on the duality transformation, since one would
guess the two allowed points are either self-dual, or dual partners.

If and how duality works for the other affine Toda theories with a boundary
will be interesting to discover. For the $ade$ series of cases the
S-matrix is self-dual, but the parameters which may be introduced at the
boundary are highly constrained \refs{\rCDRSa ,\rBCDRa}, and known
merely to be choices of
sign. Presumably, the duality transformation could relate the possibilities
in pairs, although some may be self-dual, and the Neumann condition
is likely, bearing in mind the result for sinh-Gordon,
to be related to the choice $(+)$ for all boundary parameters. The other
affine Toda models are arranged in pairs related by duality and it
is known that for most of them there are one or possibly two
extra parameters permitted at the boundary. It will certainly be
interesting
to understand how duality will work for these cases. Up to now,
some of the classical backgrounds have been found
\NRF\rBa\Ba\refs{\rCDRa ,\rBa} but there has
been no attempt to calculate the reflection factors perturbatively,
except for the Neumann condition.

Finally, one could take a different point of view.
There remains the (unwelcome?) possibility
that the weak-strong coupling transformation does not relate pairs of
boundary
conditions at all, that

$$E(a,\beta )=2a\left(1-{B\over 2}\right), \qquad F=0,$$
is the correct parametrisation, that the Dirichlet condition
and the condition
with $a=0$ are self-dual, while all the rest within the given class
have reflection factors whose large $\beta$ limit is $-1/(1)^2$.

\newsec{Acknowledgements}

The author wishes to thank the Institute of Theoretical Physics of the
University of California at Santa Barbara  for hospitality during
the period when most of this work was done, and in particular Kareljan
Schoutens and Andr\'e LeClair for comments. He also wishes to thank the
members of the Yukawa Institute of Kyoto University for discussions,
particularly Ryu Sasaki. This research was supported
in part by the National Science Foundation under Grant No. PHY94-07194,
and by a Royal Society/British Council --- Japanese Society for the
Promotion of Science Exchange Programme.

\vfill\eject
\listrefs

\end